\documentstyle[preprint,aps]{revtex}
\begin{document}
\preprint{UMP--95--65}
\title{\large\bf Analytic Solution for the Ground State Energy of the 
Extensive Many-Body Problem}  
\author{\large Lloyd C.L. Hollenberg and N.S. Witte}
\address{Research Centre for High Energy Physics,\\ School of Physics, 
University of Melbourne,\\
Parkville, Victoria 3052, AUSTRALIA.}
\maketitle
\begin{abstract}
A closed form expression for the ground state energy density
of the general extensive many-body problem is given in terms
of the Lanczos tri--diagonal form of the Hamiltonian. Given 
the general expressions of the diagonal and off-diagonal 
elements of the Hamiltonian Lanczos matrix, $\alpha_n(N)$ and 
$\beta_n(N)$, asymptotic forms $\alpha(z)$ and $\beta(z)$ 
can be defined in terms of a new parameter $z\equiv n/N$
($n$ is the Lanczos iteration and $N$ is the size of the system). 
By application of theorems on the zeros of orthogonal polynomials
we find the ground-state
energy density in the bulk limit to be given in general by 
${\cal E}_0 = {\rm inf}\,\left[\alpha(z) - 2\,\beta(z)\right]$.  
\end{abstract}
\eject
\section{Introduction}
Finding analytic means of calculating details of the energy spectrum of 
strongly interacting many-body systems has long been a goal of theoretical 
physics, with 
applications from low energy scales in condensed matter physics to high
energy particle physics. In particular, accurate calculations of the ground
state energy are important to settle issues relating to the character of
the ground state, or vacuum state, itself. While there has arisen a range
of theoretical tools to tackle problems that cannot be treated by
perturbation theory, many suffer from the effects of uncontrolled
approximation to the extent that each, alone cannot be trusted to give a
reliable answer. It is only when several independent methods point in the
same direction, that one can confidently state something about the
character of a system. Examples include mean field theories neglecting
correlations in fluctuations thereby reinforcing a tendency to order; exact
diagonalisation and Monte-Carlo studies, at zero temperature, on relatively 
small clusters and the extrapolation to the bulk limit; finite temperature 
Monte-Carlo simulation of somewhat larger systems, but at temperatures above 
the interesting low energy scales.

In this paper a simple fundamental relationship is found between the
tri-diagonal form of the Hamiltonian and the ground state energy density.
A general theorem is proved for the ground state energy density 
in terms of the coefficients generated by the Lanczos method,
evaluated in a limiting process incorporating convergence of the
Lanczos iterates and the thermodynamic limit.
The Lanczos method is based on the following recursion: 
starting from an appropriate trial state, one
recursively generates new basis states with repeated application of the
Hamiltonian
\begin{equation}
|v_{n}\rangle = {1\over \beta_{n-1}}\biggl[(H -
\alpha_{n-1})| v_{n-1}\rangle -
\beta_{n-2}| v_{n-2}\rangle\biggr],
\end{equation}
where $\alpha_n = \langle v_n|H|v_n\rangle$ and $\beta_n = 
\langle v_{n+1}|H|v_n\rangle$.
At the $n$th iteration of this recursion the Hamiltonian matrix
in the new basis of states is given by the tri-diagonal, $T_n$, i.e.
\begin{eqnarray}
H \rightarrow T_n =  
\left[
\begin{array}{ccccccccc}
\alpha_1 &  & \beta_1 &  &   &  &   &  &   \\
\beta_1 &  & \alpha_2 &  & \beta_2 &  &   &  &   \\
  &  & \beta_2 &  & \ddots &  & \ddots &  &   \\
  &  &   &  & \ddots &  & \alpha_{n-1} &  & \beta_{n-1} \\
  &  &   &  &   &  & \beta_{n-1} &  & \alpha_{n} 
\end{array}
\right].
\end{eqnarray}

The power of the Lanczos phenomenon is that the dimension of the tri-diagonal 
basis, required to describe the low-lying states of the system, 
determined by the recursion level is significantly smaller than the original
basis. 
The outermost eigenvalues of $T_n$ rapidly converge to those of the
Hamiltonian. 
In computing the Lanczos coefficients, $\alpha_n$ and $\beta_n$, 
exactly one is usually restricted to an early termination of the 
recursion in the analytic case treating large systems, 
or to small systems in a numerical calculation taken to complete convergence. 
To date all methods, however the Lanczos basis has been generated,
require the numerical diagonalisation of the tri-diagonal $T_n$ matrices
for the ground state energy. 
In this paper we will demonstrate how the diagonalisation can be carried
out analytically for extensive systems, thereby providing a solution for 
the ground state energy density in terms of the tri-diagonal form. 
By introducing a new parameter, $ z=n/N $, where $ N $ is the size 
of the system, the asymptotic forms
$\alpha(z)$ and $\beta(z)$ are related to the ground state energy by
\begin{equation}
{\cal E}_0 = {\rm inf}\,\left[\alpha(z) - 2\,\beta(z)\right] \ .
\end{equation}

\section{Thermodynamic Limit of the Tri-Diagonal Form}

Using the initial state, $|v_1\rangle$, one forms Hamiltonian moments,
and from these the connected moments,
\begin{equation}
\langle H^n \rangle_c \equiv \langle v_1 |H^n|v_1\rangle_c.
\end{equation}
The connected moments encapsulate the essential physics of the system
because they scale with the size of the system, $N$, as
\begin{equation}
\langle H^n \rangle_c \equiv c_n\,N.
\end{equation}
Here $N$ is a quantitative measure of the size of the system whether
it be the number of sites in a lattice of localised spins, or the
volume in a continuum model with itinerant particles.
Although this form is restricted to the ground state or vacuum sector of
the model, generalisations can be easily made to excited states.

A $1/N$ expansion of the Lanczos matrix elements, $\alpha_{n}(N)$ and 
$\beta_{n}(N)$, reveals a surprising analytic property -- a simple 
polynomial $n$ dependence. 
In terms of the connected coefficients $c_n$, this cluster expansion for the 
Hamiltonian $density$ are\cite{p-exp,proof} 
\begin{eqnarray}
\alpha_{n}(N)&=& c_1 + (n-1){1\over N}\,\left[{c_{3}\over c_{2}}\right]
 + (n-1)(n-2){1\over N^2}\,\left[{3 c_3^3 - 4 c_2 c_3
c_4 + c_2^2 c_5 \over 4 c_2^4}\right] + O\left({1\over N^3}\right),\nonumber\\
\\
\beta^2_{n}(N)&=& {n\over N} c_{2}
+ n(n-1){1\over N^2}\left[{c_{2}c_{4} - c_{3}^{2}\over 2 c_{2}^{2}}\right]
\nonumber\\
\nonumber\\
&+& n(n-1)(n-2){1\over N^3} \left[{21 c_{2} c_{3}^2 c_{4}-12 c_{3}^4
- 4 c_{2}^2 c_{4}^2 - 6 c_{2}^2 c_{3} c_{5} + c_{2}^3 c_{6}
\over 12 c_{2}^5}\right] + O\left({1\over N^4}\right). 
\end{eqnarray}

We first define a parameter $z\equiv n/N$ which remains finite as the
Lanczos recursion proceeds in the bulk limit, $n\rightarrow\infty$ and 
$N\rightarrow\infty$. In this limit the expansions become a series in $z$, 
i.e.
\begin{eqnarray}
\alpha(z) &\equiv& \lim_{n,N\rightarrow\infty} \alpha_n(N) =
c_1 + z\,\left[{c_{3}\over c_{2}}\right]
 + z^2\,\left[{3 c_3^3 - 4 c_2 c_3
c_4 + c_2^2 c_5 \over 4 c_2^4}\right] + O(z^3),\nonumber\\
\\
\beta^2(z)^&\equiv& \lim_{n,N\rightarrow\infty} \beta^2_n(N) =
z\,c_{2} + z^2\left[{c_{2}c_{4} - c_{3}^{2}\over 2 c_{2}^{2}}\right]
\nonumber\\
\nonumber\\
&+& z^3 \left[{21 c_{2} c_{3}^2 c_{4}-12 c_{3}^4 
- 4 c_{2}^2 c_{4}^2 - 6 c_{2}^2 c_{3} c_{5} + c_{2}^3 c_{6}
\over 12 c_{2}^5}\right] + O(z^4). 
\end{eqnarray}

The cluster expansion guides us to the observation that, more generally,
we have for the exact problem the asymptotic forms in the $n\rightarrow\infty$
and $N\rightarrow\infty$ regime 
\begin{eqnarray}
\alpha_n(N) &=& \alpha(z) + O(1/N),\nonumber\\
\beta^2_n(N) &=& \beta^2(z) + O(1/N).
\end{eqnarray} 
We observe here a confluence of the two limiting regimes: that of convergence
of the Lanczos iterates and of the thermodynamic limit into a single
scaled Lanczos iteration number $z$.

\section{Orthogonal Polynomials and Van Doorn's Theorem}

The connection between the Lanczos tri-diagonal form of the
Hamiltonian and the associated system of orthogonal polynomials is 
simple. The characteristic polynomial $D_n(x) = 
{\rm det}(T_n - xI_n)$ of the Lanczos tri-diagonal matrix representation 
satisfies the following recursion relation:
\begin{eqnarray}
D_n(x) = (\alpha_n - x) D_{n-1}(x) - \beta^2_{n-1} D_{n-2}(x),
\end{eqnarray}
which in turn defines $P_n(x) \equiv (-1)^nD_n(x)$ as an orthogonal 
polynomial, the zeros of which are the eigenvalues of
the $T_n$ matrices. 
The analytic forms for the $\alpha_n(N)$ and $\beta_n(N)$
define a special class of orthogonal polynomials relevant to the many-body 
problem which are distinguished by a certain dependence on the size
parameter $N$.

For the orthogonal polynomials $P_n(x)$ satisfying the recursion relation 
\begin{eqnarray}
P_n(x) = (x - \alpha_n) P_{n-1}(x) - \beta^2_{n-1} P_{n-2}(x)
\end{eqnarray}
there exits a powerful theorem by Van Doorn\cite{doorn} on the lower bound 
on the lowest zero which has been generalised by Ismail and
Li\cite{ismail} to include an upper bound to the largest zero.
Simply stated Ismail and Li's result is the following:
if $x_k^{(1)}$ and $x_k^{(k)}$ are the smallest and 
largest zeros respectively of $P_k(x)$ with $k > 1$ then they are bounded 
by the interval $(A,B)$ where  
\begin{eqnarray}
A &= {\rm min}\{f^{-}_n : 1 \le n < k\}\nonumber\\
B &= {\rm max}\{f^{+}_n : 1 \le n < k\}
\end{eqnarray} 
and the bound sequence is given by 
\begin{eqnarray}
f^{\pm}_n = {1\over 2}\left[ (\alpha_n + \alpha_{n+1})
            \pm \sqrt{(\alpha_n - \alpha_{n+1})^2 
          + {4\over a_n} \beta_n^2}\quad\right].
\end{eqnarray}
In the above expression for $f^{\pm}_n$, $\{a_n\}_1^{k-1}$ is a chain sequence.
That is, there exists a parameter sequence $\{g_n\}_1^k$ for which we have 
the factorisation of $a_n$
\begin{eqnarray}
a_n = g_{n+1} (1 - g_n),  \qquad 1 \le n < k
\end{eqnarray}
where $0\le g_1<1$ and $0<g_n<1$ for $1 < n \le k$. 
Van Doorn's theorem is for strict equality, whereby
maximising the lower bound with respect to the parameter
sequence gives the lowest eigenvalue exactly:
\begin{equation}
x_k^{(1)} = 
\max_{\{g\}}\,\left[\min_n \{f^{-}_n : 1 \le n < k\}\right]
\end{equation}
for $k > 1$.
Only a technical difference occurs with this theorem in that 
$g_1=0$ and $g_k=1$.

For finite $N$ the termination of the Lanczos recursion occurs at some
$n_{\rm max}$ when the sector of Hilbert space, determined by the trial
state, has been exhausted. In general the basis of states grows 
faster than any linear enumeration with $N$. Taking the minimum of $f^{-}_n$
to occur at some ${\bar n} = {\bar z}N$ (${\bar n} < n_{\rm max}$), 
the asymptotic forms for $\alpha_n(N)$ and $\beta_n(N)$ give the 
leading order behaviour of the lowest energy level as 
\begin{equation}
x_{n_{\rm max}}^{(1)} = 
\max_{\{g\}}\,\,\left[\alpha({\bar z}) - {1\over \sqrt{a_{{\bar n}}}}
\,\beta({\bar z}) + O(1/N)\right].
\end{equation}

Since $\beta_n(N)>0$, to maximise the RHS of the above expression with 
respect to the chain sequence we may choose the maximal constant chain 
sequence\cite{ismail}
\begin{equation}
a = {1\over {4 \,{\rm cos}^{2}\left(\pi \over n_{\rm max} + 2\right)}}.
\end{equation}
We can now obtain the ground state energy density, ${\cal E}_0$, in the bulk 
limit:
\begin{equation}
{\cal E}_0 = \lim_{N\rightarrow\infty}\,x_{n_{\rm max}}^{(1)}.
\end{equation}

In the $N\rightarrow\infty$ limit, we also have $n_{\rm max}\rightarrow\infty$ 
giving $a\rightarrow 1/4$, and it is straightforward to establish that 
in general ${\bar z}$ is obtained by finding the greatest lower bound with 
respect to $z$. The ground state energy density of the general extensive 
many-body problem in the bulk limit is therefore 
\begin{equation}
{\cal E}_0 = \inf_{z>0}\,\left[\alpha(z) - 2\,\beta(z)\right].
\end{equation}

\section{Examples}

As a first illustration of this exact analytic diagonalisation of extensive
systems we consider the purely mathematical model defined by the tri-diagonal
form 
\begin{eqnarray}
\alpha_n(N) &=& \left(1 + {a\,n\over N^2}\right)^N,\nonumber\\
\beta_n(N) &=& \left(1 + {b\,n\over N^2}\right)^N - 1.
\end{eqnarray}
The lowest eigenvalue in the $N\rightarrow\infty$ limit is
\[
{\cal E}_0 = {\rm inf}\,\left[e^{a z} - 2 e^{b z} + 2\right]
= \left({r\over 2}\right)^{r/(1-r)}\,\left(1-r\right) + 2,
\]
where $r = a/b$. The minimum occurs at $z_{\rm min} = {\rm ln}(2/r)/b(r-1)$. 
It is a straightforward matter to numerically diagonalise this system for
increasing $N$ to demonstrate the convergence of the numerical results to the 
analytic expression. 
The results of this exercise are shown in Figure 1 (for a
typical case $a=3$ and $b=2$) where we
plot the error, defined as the difference from the exact value, of
numerical diagonalisations for increasing $N$. The approach to the 
analytic solution is clear as $N\rightarrow\infty$. 

Secondly we compare the 
the analytic diagonalisation with numerical results using the 
truncated Lanczos coefficients given by the plaquette
expansion. We do this here for the case of the anti-ferromagnetic
Heisenberg chain, for which the expansions have been derived to high
order using the classical ${\rm N{\acute e}el}$ state as the trial state. 
This state is an poor choice for the isotropic Heisenberg model but is
sufficiently simple to allow the computation of moments up to
$< H^{28} >_c$. 
Employing the traditional analysis of the plaquette expansion the $T_n$
matrices are diagonalised numerically for increasing chain size 
$N$\cite{1DAFH}. In Figures 2 and 3 we demonstrate the approach of the 
numerical diagonalisation for the ground state to the analytic expression as 
the $N\rightarrow\infty$ limit is approached. We show the two typical cases
which can occur due to the truncation of the expansions for $\alpha(z)$ 
and $\beta(z)$. Figure 2 corresponds to an order of the plaquette
expansion for which $\alpha(z) - 2\,\beta(z)$ has no minimum -- a point of 
inflection develops, reflecting the fact that the expansion naturally breaks 
down at some large enough value of $z$. Figure 3 shows
a case where a minimum develops -- the numerical values for the
lowest eigenvalue in the large volume limit match onto the value of
at the minimum. 
 
We must emphasis that the plaquette expansion is a series expansion of
the exact Lanczos coefficients about $z=0$ and truncated, and is therefore
only reliable for small $z$. The error involved grows rapidly with $z$ which 
can occur with either sign, and thus the large $z$ behaviour of the
truncated approximations bear no relation to the exact behaviour.
Because of this we don't expect that at every order at which the truncation
is made, that a minima would arise. 

\section{Conclusions}

We have found an expression for the ground state energy density for the 
extensive many-body problem which is completely general and,
since the Hamiltonian was diagonalised exactly, is non-perturbative. 
Given the Hamiltonian in tri-diagonal form the expression can be used
immediately. So far, the exact analytic transformation of a 
system to tri-diagonal form has not been achieved for any
examples of solvable systems. 
This state of affairs may change as the theorem proved here gives impetus
to efforts in this direction.
However, an immediate approximate tri-diagonalisation of the
general problem does exist in the plaquette expansion and studies of the
usefulness of this method with better trial states also merits further 
work. The analytic
nature of this expansion uncovered the existence of the scaled Lanczos
iteration parameter 
$z=n/N$ and the asymptotic forms $\alpha(z)$ and $\beta(z)$ for the
exact problem.

An interesting extension of this work is to consider a similar analysis
on excited states. An indication that the solution for the mass-gap of the 
general extensive many-body system may in fact be possible is the fact 
that the mass-gap of the first order plaquette expansion has already been  
solved analytically\cite{first_order,massgap}. 
Such a generalisation is presently under active investigation.
\eject

\begin{figure}[htb]
\caption{Comparison of the analytic solution with numerical
diagonalisation for the simple case, 
$\alpha_n(N) = (1 + a n/N^2)^N$ and 
$\beta_n(N) = (1 + b n/N^2)^N - 1$, with $a=3$ and $b=2$.}
\end{figure}

\begin{figure}[htb]
\caption{Plaquette expansion of the 1D AFH model at an order ($1/N^{11}$)
where a point of inflection develops due to the breakdown of the
expansion. The numerical diagonalisation for increasing $N$ converges to the
function $\alpha(z)-2\beta(z)$.}
\end{figure}

\begin{figure}[htb]
\caption{Plaquette expansion of the 1D AFH model at an order ($1/N^{7}$)
where a minimum develops in $\alpha(z)-2\beta(z)$. The numerical 
diagonalisation data at $N=5\times10^4$ clearly display the convergence to 
the value of ${\rm min}[\alpha(z)-2\beta(z)]$.}
\end{figure}

\eject

\vspace*{30mm}
\begin{center}
\bf Acknowledgements
\end{center}
This work was supported by the Australian Research Council. Preliminary
numerical work by J. Haskian is gratefully acknowledged. 

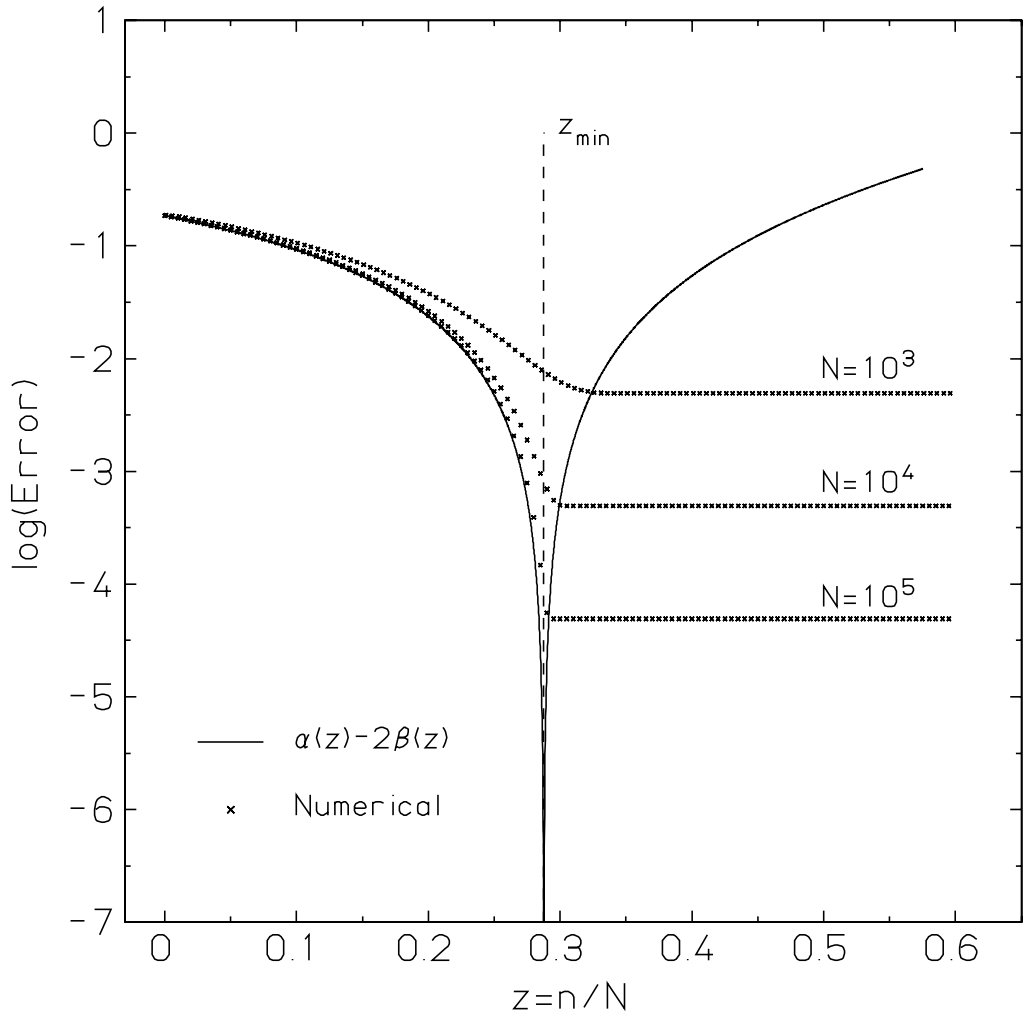
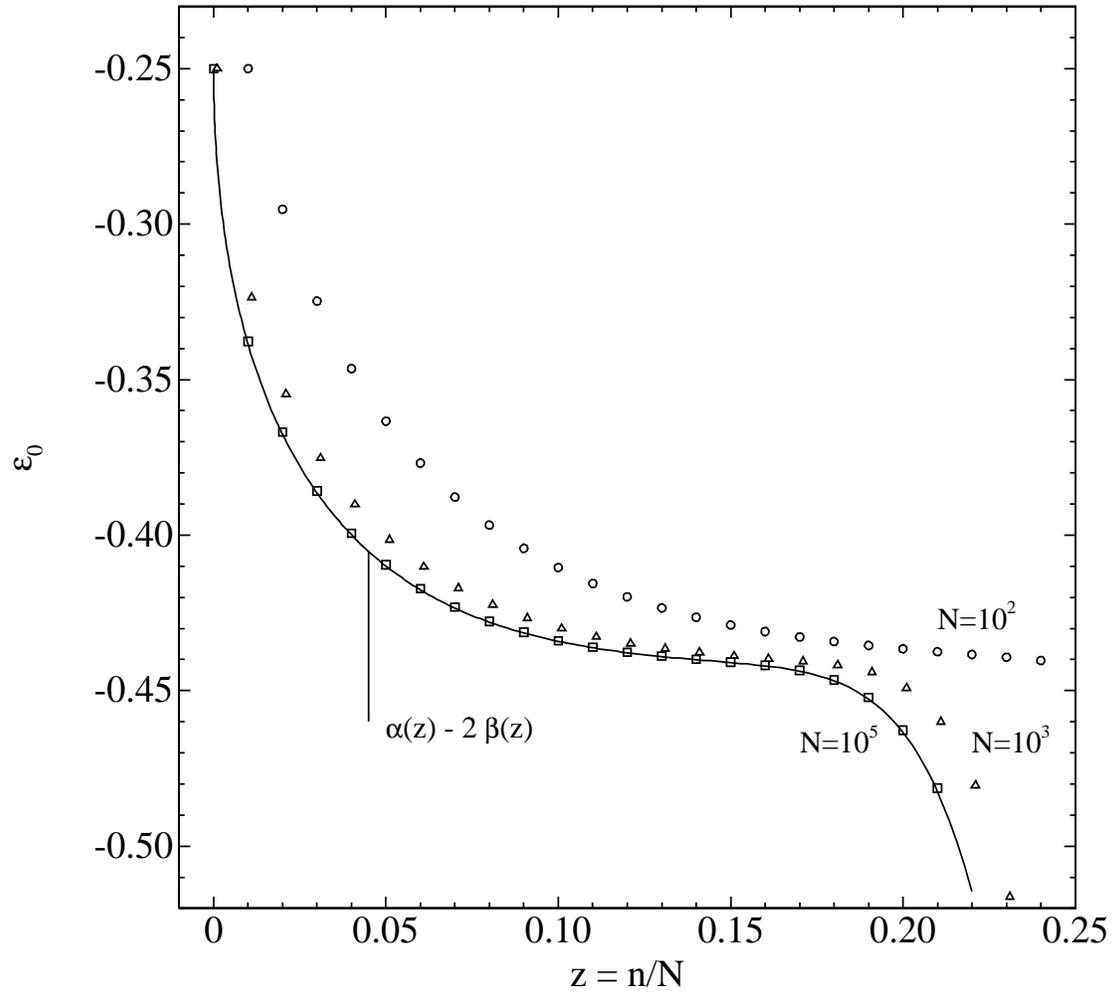
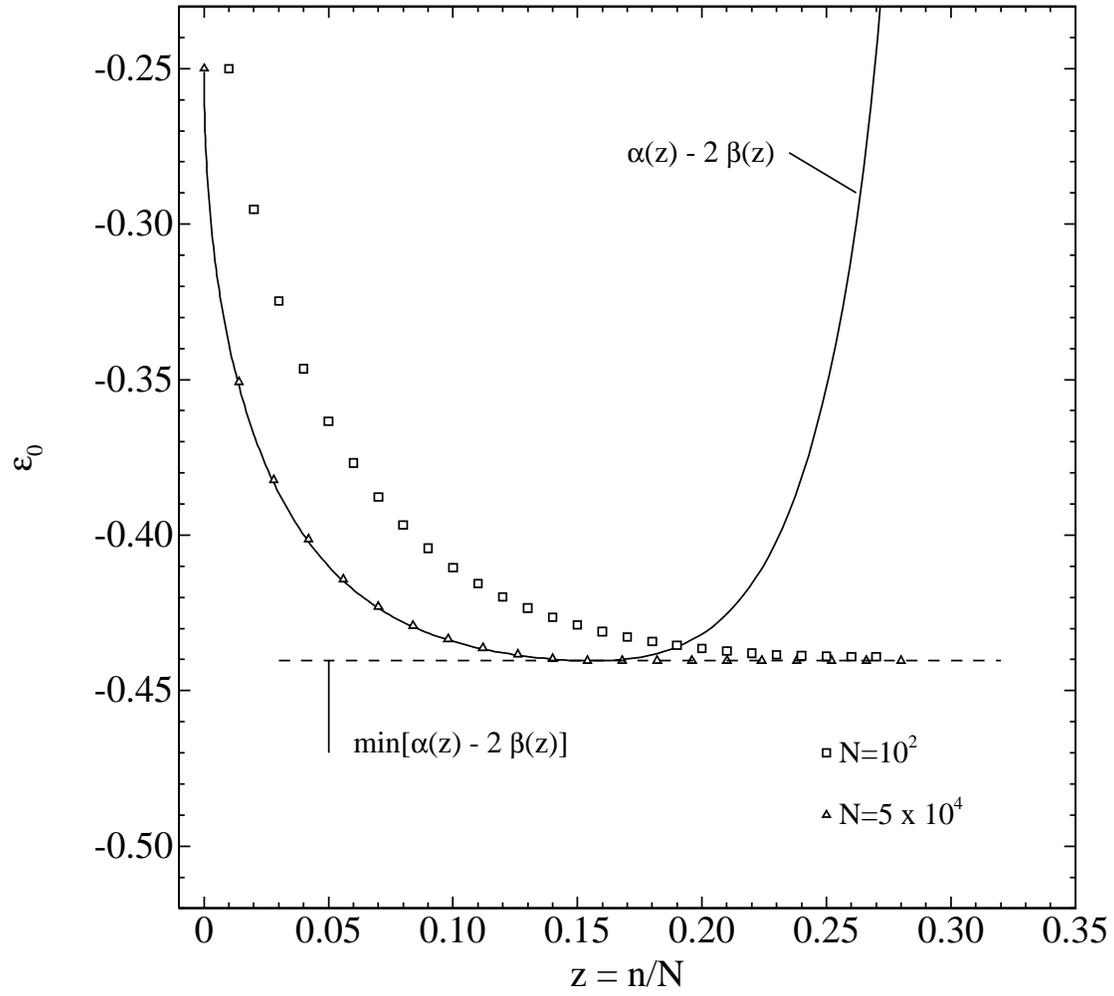
\end{document}